\documentclass[12pt, preprint]{aastex}

\shorttitle{Holography of Simulated Solar Convection}
\shortauthors{Braun and Birch}

\begin{document}

\title{Helioseismic Holography of Simulated Solar Convection and 
Prospects for the Detection of Small-Scale Subsurface Flows}

\author{D.~C.\ Braun, A.~C.\ Birch}
\affil{NorthWest Research Associates, CoRA Division, 3380 Mitchell Lane, 
Boulder, CO 80301, USA}
\email{dbraun@cora.nwra.com aaronb@cora.nwra.com}
\author{D.\ Benson\altaffilmark{1}, R.~F.\ Stein}
\affil{Physics and Astronomy Department, Michigan State University, 
East Lansing, MI 48824, USA}
\email{dbenson@kettering.edu stein@pa.msu.edu}
\and
\author{\AA.\ Nordlund}
\affil{Niels Bohr Institute, Copenhagen University, 
Juliane Maries Vej 30, DK-2100 Copenhagen \O, Denmark} 
\email{aake@astro.ku.dk}
\altaffiltext{1}{present address: Department of Mechanical 
Engineering, Kettering University, Flint, MI 48504, USA}

\begin{abstract}
We perform helioseismic holography on realistic solar convection 
simulations and compare the observed travel-time perturbations 
with the expected travel times from the horizontal flows in the 
simulations computed from forward models under the assumption of 
the Born approximation. We demonstrate reasonable agreement 
between the observed and model travel times which reinforces the 
validity of helioseismic holography in the detection of subsurface 
horizontal flows.  An assessment is made of the uncertainty of 
the measured $p$-mode travel-times, from the root-mean-square of 
the residuals.  From the variation of the signal-to-noise ratio 
with depth, we conclude that the helioseismic detection of 
individual flow structures with spatial scales of 
supergranulation or smaller is not possible for depths below about 
5 Mm below the surface over time scales less than a day. The 
travel-time noise estimated from these simulations appears to be 
similar to noise in measurements made using solar observations. 
We therefore suggest that similar limitations exist regarding the 
detection of analogous subsurface flows in the Sun. A study of 
the depth dependence of the contribution to the travel-time 
perturbations for focus depths between 3 and 7 Mm is made, showing 
that approximately half of the observed signal originates within 
the first 2 Mm below the surface. A consequence of this is a rapid 
decrease (and reversal in some cases) of the travel-time 
perturbations with depth due to the contribution to the measurements 
of oppositely directed surface flows in neighboring convective cells. 
This confirms an earlier interpretation of similar effects reported 
from observations of supergranulation.

\end{abstract}

\keywords{Sun: helioseismology, interior}

\section{Introduction}

The understanding of solar interior dynamics, as manifested
in mass flows with a variety of temporal and spatial scales beneath the
surface of the Sun, has been considerably advanced by 
both global- and local-helioseismic observations. 
Some recent reviews of the progress in helioseismology of flows
(and other properties of the solar interior) include 
\citet{Jcd2002}, \citet{Thompson2003}, \citet{Gizon2005},
\citet{Miesch2005}, and \citet{Thompson2006}. In addition, progress has been made
with increasingly improving numerical modeling procedures and 
computational resources. Examples of numerical models and simulations
of wave propagation relevant to helioseismic studies 
include  \citet{Birch2001}, 
\citet{Jensen2003}, \citet{Tong2003}, \citet{Mansour2004},
\citet{Benson2006}, \citet{Hanasoge2006}, \citet{Khomenko2006}, 
\citet{Shelyag2006}, 
\citet{Parchevsky2007}, \citet{Shelyag2007}, and \citet{Zhao2007} 
In addition to providing physical insights into the origin
and evolution of flows inferred through helioseismic observations
and analysis, numerical models such as those cited above are 
being used to validate the observational and analysis tools
themselves. 
The application of analysis methods to
numerical (or ``artificial'') data can potentially lead to 
improvement in both theory and observations of the phenomena
being explored \citep[e.g.][]{Werne2004}.

The mean solar rotation, and its variation with
depth, latitude, and time, is detectable by both
global and local helioseismic techniques, and comparisons between
methods provides checks on the reliability of the analysis and
modeling \citep[e.g.][]{Giles1998, Basu1999, Basu2000, 
Haber2000, Haber2002, Zhao2004}. Confidence in local helioseismic
methods is also obtained by comparing 
models, derived using different procedures and assumptions, of
low-amplitude flows such 
meridional circulation 
\citep[e.g.][]{Giles1997, Giles1998, Braun1998, Haber2002, Hughes2003, Zhao2004,
Chou2005, Gonzalez2006, Svanda2007}. Intercomparisons of methods and datasets
can sometimes lead to the discovery of important systematic effects or artifacts
\citep{Gonzalez2006}.

The helioseismic study of ``small-scale'' structures within the
Sun, defined here by wave-speed, flow, or magnetic perturbations in
the solar interior with spatial scales less than a few tens of
Mm is a particularly challenging pursuit.  
This is because the structures
being probed have sizes which are of the order of, or perhaps only a
few times larger than, the wavelength of the $p$- or $f$-modes
used in the analysis. In this work we consider small-scale flows 
representative of solar supergranulation.  The effects on mode
frequencies and travel times of acoustic waves caused by
even smaller-scale flows (e.g.\ granulation) have also been explored
\citep[e.g][]{Murawski1993a, Murawski1993b, Petrovay2007}, but
are not considered here. Our simulations do not include magnetic
fields and consequently we are unable to assess
the effects of magnetic fields on the modeling 
or interpretation of helioseismic measurements of these flows. 
This important issue, however, continues to be extensively explored
in the context of global helioseismology, asteroseismology,
and local helioseismology \citep[e.g.][]{Bogdan2000, Jcd2002,
Gizon2005, Thompson2006}.

Supergranulation has been a focus of local helioseismology
for over a decade \citep[e.g.][]{Duvall1997, Kosovichev1997,
Duvall2000, Gizon2000, Braun2003, Gizon2003, Zhao2003, Braun2004}.
However, as noted by \citet{Gizon2005} there is no definitive
consensus on even the depth of the supergranulation phenomenon. 
Some inversions of $p$-mode travel times indicate the presence of a
``return flow'' (oppositely directed flows from what is observed at
the solar surface) at various depths, but typically $\sim$ 10 Mm, below
the photosphere \citep{Duvall1997, Zhao2003}. \citet{Braun2004} have suggested
that an observed change in sign in travel-time perturbations with increasing depth
of the penetration of the modes may represent a contamination of
the signal from neighboring supergranules. How this might effect
the inversions, however, is not clear. \citet{Zhao2007}
find that travel-time inversions on simulated supergranular-sized
convection do not
reproduce the simulation flow fields at depths below the photosphere
greater than 3 Mm. How deep we are able to
reliably measure subsurface flows due to supergranulation and 
other small-scale patterns remains an important, but unanswered,
question.

In this paper, we explore the prospects
for helioseismic probing of small-scale flows in the
solar interior by applying helioseismic holography
\citep{Lindsey1997}
to recent numerical simulations of solar convection.
To do this we compare the ``observed'' signatures of flows (in this
case, maps of the perturbations to $p$-mode 
travel times) obtained by applying helioseismic holography
to the surface of realistic solar simulations, to the expected
(''model'') signatures obtained directly from the simulated flows.
This comparison facilitates an assessment of both the expected
helioseismic signals and the uncertainties in the
observations (i.e.\ departure of
the observed signatures from the expected values). 
While the degree of similarity between observed
and model signatures provides a critical validation of 
our specific helioseismic analysis our principle goal is 
more general than a test of a single method of observation
or modeling procedure.
In particular, the observational uncertainties estimated from
this {\it forward-modeling} experiment are used to assess the general prospects
for the seismic detection of small-scale flows in the solar interior. 
This is possible to the extent that 1) our simulation measurements 
have uncertainties characteristic
of (or at least no worse than) local helioseismic observations 
in general, and 2) flows 
in the solar interior have properties similar to  
those in the simulations.  The validity of the 
second condition is maximized by using realistic solar convection
simulations \citep{Benson2006} as described
in \S~\ref{sims}. 
To ensure the validity of the first of these conditions, we 
employ helioseismic
holography in the ``lateral vantage'', which (as explained below) is
designed to utilize most of the $p$-modes propagating through a
specific target point (focus) in the solar interior.  We also 
compare the uncertainties estimated in the simulation observations 
with results obtained from the Michelson Doppler Imager (MDI) onboard
the {\it Solar and Heliospheric Observatory (SOHO).}
Our analysis
and forward modeling procedures are described in 
more detail in \S~\ref{analysis} and \S~\ref{models} respectively.  
The results of the comparison are
shown in \S~\ref{results}, followed in \S~\ref{discussion} by a discussion of
the implications of these results.

\section{Simulations} \label{sims}

The 3D simulations of solar convection on supergranular scales 
employed in this study were performed and described 
by \cite{Benson2006} and have already been used to validate 
surface and subsurface flow diagnostics \citep{Georgobiani2007, Zhao2007}.
The simulations are carried out in 
a box 48 by 48 Mm horizontally and 20 Mm deep.
The conservation equations for density,
momentum and internal energy are solved in conservative form, on a
three-dimensional staggered mesh, using sixth order finite difference
spatial derivatives and a low memory, third order Runge-Kutta time
advance.  The grid is uniform in horizontal directions and stretched
in the vertical (stratified) direction.  Horizontal boundary
conditions are periodic, while top and bottom boundary conditions
are open.  Inflows at the bottom boundary have constant pressure,
specified entropy and damped horizontal velocities.  Outflow boundary
values are obtained by extrapolation.  The code uses a tabular
equation of state, that includes local thermodynamic equilibrium
(LTE) ionization of the abundant elements as well as hydrogen
molecule formation, to obtain the pressure and temperature as a
function of log density and internal energy per unit mass using the
Uppsala atmosphere package \citep{Gustafsson1975}.  
Radiative heating/cooling is determined by solving the
radiation transfer equation in both continua and lines using the
Feautrier method, assuming Local Thermodynamic Equilibrium 
\citep{Nordlund1982, Stein2003}.

The acoustic waves ($p$-modes) generated by convective motions
in these simulations have properties remarkably similar to
observed solar $p$-modes \citep{Georgobiani2007}. One notable
exception is an enhancement of acoustic power at higher temporal 
frequencies relative to observed solar values.
A fortunate consequence of this is the ability to perform 
and test helioseismic procedures with modes extending to higher
frequencies than normally used with solar data. 

\begin{figure}[htbp]
\epsscale{1.0}
\plotone{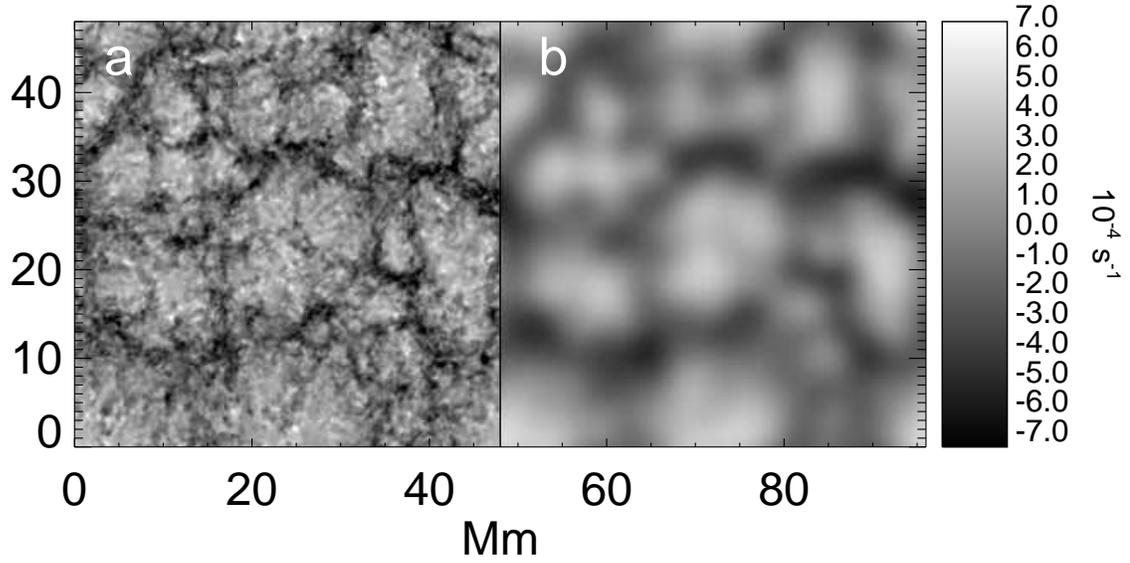}
\caption{a) The horizontal divergence of the time average of the 
simulated flows at a depth of 4 Mm below the surface. 
Positive values of the divergence, identified by 
bright regions, correspond to outflows. 
b) the same flow divergence smeared with a two
dimensional Gaussian function with a FWHM of 4 Mm.}
\label{cells}
\end{figure}

While power spectra of the convective motions in these
simulations demonstrate the presence of structures
over a broad range of spatial scales \citep{Benson2006, Georgobiani2007},
we expect that local helioseismic techniques are
best suited for probing structures larger than the
wavelength of the $p$-modes employed in the analysis.
Figure~\ref{cells}a  shows the horizontal 
divergence of the 8.53 hr time-average 
of the simulated flows at a depth of 4 Mm below the surface. 
Figure ~\ref{cells}b shows the same flow divergence after applying
a Gaussian smoothing with a full-width-half-maximum (FWHM) of 4 Mm.
This width is of the order of the wavelength of
$p$-modes with a temporal frequency
of 3 mHz just below the photosphere 
(or equivalently, the wavelength at 5mHz of modes at a depth 
of about 4 Mm).
Evident in both panels (but especially in the smoothed panel)
are cellular structures, the most prominent having diameters
on the order of 20 Mm. Presumably these structures are
the most amenable to helioseismic detection with mode
wavelengths greater than 4 Mm, and serve as a proxy for
solar supergranulation which has a somewhat
larger scale ($\sim 30$ Mm). The simulated flow structures are
coherent to depths of about 12 Mm, below which appear
weak counterflows (see Figure ~\ref{ratio}).

We use 8.53 hours of the simulations in our analysis, which
is typical of solar helioseismic observations and is less than
the lifetimes of individual solar supergranules 
($\sim 25$ hr).  
We apply helioseismic holography (see \S~\ref{analysis}) 
to the vertical component of
velocity sampled a height of 200 km above the 
continuum optical depth of unity in the background
stratification (hereafter referred to as
the ``surface'' of the simulations). We use flows
sampled only every 30 seconds in time, yielding 1024 total time
samples.  The original simulations are computed in a 500 by 500 horizontal
grid (giving a horizontal resolution of 96 km) and are for
our purposes degraded in resolution by a factor of two in each 
direction by pixel averaging.

\section{Analysis} \label{analysis}

Helioseismic holography (hereafter HH) is a method based on
the phase-coherent imaging of the solar interior acoustic field.
It computationally extrapolates the surface acoustic field
into the solar interior \citep{Lindsey1997, Lindsey2000}
to estimate the amplitudes of the waves propagating into
and out of a focus point at a chosen depth and position in the solar
interior. These amplitudes, called the ingression and
egression, are estimated by a convolution of the surface oscillation signal
with appropriate Green's functions \citep{Lindsey2000}.
Here, HH is performed in the wavenumber-frequency (Fourier)
domain using a set of fixed frequency bandpasses with a width of 1 mHz
and centered at frequencies, ${\nu}_0$, of 3, 4, 5, and 6 mHz.
While previous applications of HH to study solar flows 
have been limited, by signal-to-noise considerations, 
to frequencies equal to and below 5.5 mHz
\citep[e.g.][]{Braun2003, Braun2004}, we are motivated by the
presence of high frequency waves in the simulations to include
a 6mHz bandpass in our analysis here. 

\begin{figure}[htbp]
\epsscale{0.8}
\plotone{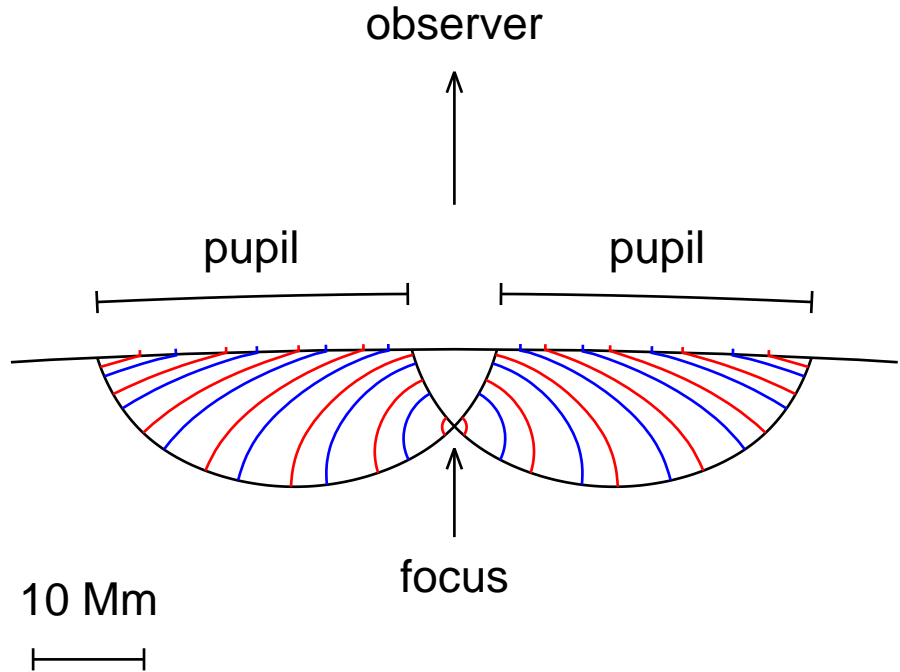}
\caption{Lateral-vantage holography with the focus 
placed at a depth of 7 Mm below
the surface. Solid black curves which pass 
through the focus at angles inclined
$\pm 45^\circ$ from the horizontal direction 
indicate the propagation of acoustic rays
which, when reaching the surface (horizontal line)
define the inner and outer radii of the pupil. The colored
curves indicate wavefronts (separated by a wavelength) 
of sample components of the
egression (red) and ingression (blue) for this focus depth.
The wavelength of the 
wave components shown here is 5.9 Mm, corresponding to
$p$-modes with a temporal frequency of 5 mHz.
If set in motion, the red wavefronts would diverge
away from the focus, while the blue wavefronts would
converge towards the focus.
}
\label{latvant}
\end{figure}

The HH analysis is performed 
in the wave-mechanical formulation \citep[][]{Lindsey2004}.
The method employed for horizontal flow diagnostics is based on the
egressions and ingressions computed in the {\it lateral
vantage} \citep{Lindsey2004} employing pupils spanning 4 quadrants
extending in different directions
(east, west, north and south) from the focus \citep{Braun2004}.
In the lateral vantage, the $p$-modes sampled by the pupil
propagate through the focal point in directions inclined up
to $\pm 45^\circ$ from the direction parallel to the
surface (Figure~\ref{latvant}).
The antisymmetric
phase shift, the difference in the phase perturbation
of waves traveling from one pupil to its opposite and the
phase perturbation of waves traveling in the reverse direction,
is sensitive to horizontal flows near the focus.
In general,
the phase perturbation $\delta\phi^{A}$
is related to an equivalent
travel-time perturbation by $\delta\tau = \delta\phi^{A} / 2\pi{\nu}_0$.
The travel time perturbations, $\delta{\tau}_x$ and $\delta{\tau}_y$
derived from antisymmetric phase shifts computed between the east/west
and north/south quadrant pairs, respectively,
provide the HH signatures sensitive to the two components of
the horizontal flow.
The sign of the travel-time perturbations are such that a 
positive velocity component will produce a negative
value of $\delta{\tau}$ (i.e.\ a reduction in the 
mode propagation time).
The lateral-vantage geometry samples  
more than 70\% of the wave modes which pass through
the focus. The remaining waves, propagating more vertically than
the waves appearing in the pupil, are substantially less sensitive to 
horizontal flows.

Table~\ref{tbl-1} lists the focus depths and the pupil 
radii used in lateral-vantage HH. The pupil radii are
defined from ray theory. The range
of (spherical-harmonic) mode degrees ($\ell$) at
4 mHz, selected
by each pupil, is listed in the table. The lower
$\ell$ value denotes the modes propagating at
$\pm 45^\circ$ 
from the horizontal direction which propagate through the
focus and reach the surface at either the inner or
outer pupil radius. The highest $\ell$ value listed
indicates modes propagating horizontally through the
focus. The mode degrees selected by 
the other frequency bandpasses scale approximately with
${\nu}_0$.  Dispersion
effects predict slight variations in the pupil radii with
wave frequency. However, experimentation 
shows that the measured travel-times do not vary 
with similar adjustments of the pupil radii 
sufficiently to affect the comparisons
and conclusions discussed here. Consequently, for convenience
and simplicity, we use the same pupil for all frequency
bandpasses for a given depth focus.

Comparisons of travel-time maps made with and without 
the use of phase-speed filters show that these 
filters are of marginal utility when applied to 
data subsequently sampled over wide pupils of
the sort employed in lateral-vantage HH.
We note that this is not the case for much of time-distance
helioseismology \citep[e.g.][]{Duvall1997, Zhao2001,  Zhao2007} or
for HH performed with narrow pupils \citep{Braun2006}, where
the use of phase-speed filters appears to provide 
a dramatic improvement in the measurements.
While a reduction in scatter is sometimes evident with  
the use of these filters in lateral vantage HH, 
we find that, for both solar data and the present simulated data, 
the effect is comparable to what can be achieved by
spatially smearing the unfiltered travel-time perturbation 
maps.  In fact, an understanding of the consequences of 
smearing the travel-time maps 
is critical for interpreting our
comparisons between observed and modeled results.
With these considerations in mind, our strategy is to
use observed travel-time maps made without 
phase-speed filters, and explore 
explicitly the direct consequences to our comparisons of
spatial averaging of the unfiltered maps by varying amounts.

The focus depths chosen to perform the ``observations''
extend down to about 8 Mm below the simulation surface.
At focus depths larger than this, egression-ingression correlations cease
to exhibit meaningful results, even though reliable correlations are
routinely observed at these depths in the Sun. This failure may be
due to the lower boundary of the simulations at 20 Mm. Even though
the lower boundary of the simulation is twice the depth of the focus 
depth where problems appear to arise, we note that the lateral-vantage 
geometry samples $p$-modes which penetrate a considerable depth
below the focus, and any reflection or absorption of these waves
at the lower boundary will likely adversely affect our observations.
The horizontal periodic 
boundaries used in the simulations ensure that HH travel-time
measurements, performed in the Fourier domain,
are usable over the full horizontal extent of the simulations. 
Thus, the data is ``infinite'' (but periodic) horizontally, and 
travel-time perturbation maps are free from
any detrimental effects caused by an abrupt termination of data at the 
edges of the domain, such as typically experienced with helioseismic
observations of the Sun.

\section{Forward Models} \label{models}
 
In this section we use the Born approximation approach 
of \citet{Gizon2002} to estimate the HH signatures that
would be expected from the flows in the simulations.
The range of validity of the Born approximation
for three-dimensional time-dependent flows in 
stratified models is not known.   

\citet{Birch2004a} studied the validity of the Born approximation as applied
to time-distance helioseismology of simple steady flows in 2d unstratified
models.  In these models, the Born approximation is typically seen to be valid
when travel-time shifts are a small fraction of the wave period. The validity
of the Born approximation for scattering from steady sound-speed
inhomogeneities has been studied extensively in the context of earth seismology
and helioseismology \citep[e.g.][]{Fan1995, Hung2000, Birch2001, Baig2003}.
In these studies, it is again typical that the Born approximation
is valid when travel-time shifts are small compared to the mean wave period
(though it can sometimes be valid well past this limit).  The extension of this
general conclusion to the current work, which involves strong time-dependent
small-scale flows, is not at all certain.

Comparison between the Born approximation estimates described
in this section and the actual travel times (\S~\ref{results})
will provide an important estimate of the validity 
of the Born approximation for complicated solar-like
flows.  In particular, we will be able to study the 
extent to which the HH signatures can be predicted
using only the time average of the flow.

As described in detail by \citet{Gizon2002} the computation
of the linear sensitivities of local helioseismic measurements
requires, in general, two steps.  The first step is the computation
of the  linear sensitivity of the measurement procedure
(e.g.\ the measurement of travel times) to small changes in 
covariance of the wavefield. The second step is the computation of
the change in the wavefield covariance introduced by small changes in the model of the solar interior (e.g. the introduction of flows or sound-speed variations). This general procedure has been applied by \citet{Birch2007a} to 
the case of time-distance measurements of flows and by \citet{Birch2007b} 
to the case of ring-diagram measurements of flows.

The application of the \citet{Gizon2002} recipe to the case of HH
 is relatively straightforward.  The result is a set of linear sensitivity
 functions (kernels), ${\bf K}$, which give the linear sensitivity of the
 anti-symmetric phase $\phi^{\rm A}$ to small amplitude steady  flows
 ${\bf v}({\bf x})$,
\begin{equation}
\delta\phi^{\rm A}= \int_\odot {\rm d}{\bf x} \; {\bf K}(\bf x) \cdot {\bf v}({\bf x}), \label{eq:kernel}
\end{equation}
where ${\bf x}$ denotes three-dimensional position in the model
and the integral  is taken over the entire three-dimensional domain 
of the simulation.
In the computation of the kernel functions we use
the normal-mode Greens functions, source model, and
damping model described by \citet{Birch2004b}.  The background
stratification is given by Model S \citep{Jcd1996}

One limitation of the normal-mode Green's functions of
\citet{Birch2004b} is that they employ the upper boundary condition that Lagrangian pressure perturbation vanishes
at the top of Model S.  As discussed by \citet{Birch2004b}, this boundary condition is reasonable for waves at frequencies
well below the acoustic cutoff frequency.  For frequencies near the
acoustic cutoff frequency, this boundary condition is likely not
appropriate.  As a result, the kernels described in this section
are not intended to model HH measurements at or above five mHz.

\section{Results}                   \label{results}

\subsection{Frequency and Depth Variations of Observed Travel-Time 
Perturbations} \label{obs}

\begin{figure}[htbp]
\epsscale{1.0}
\plotone{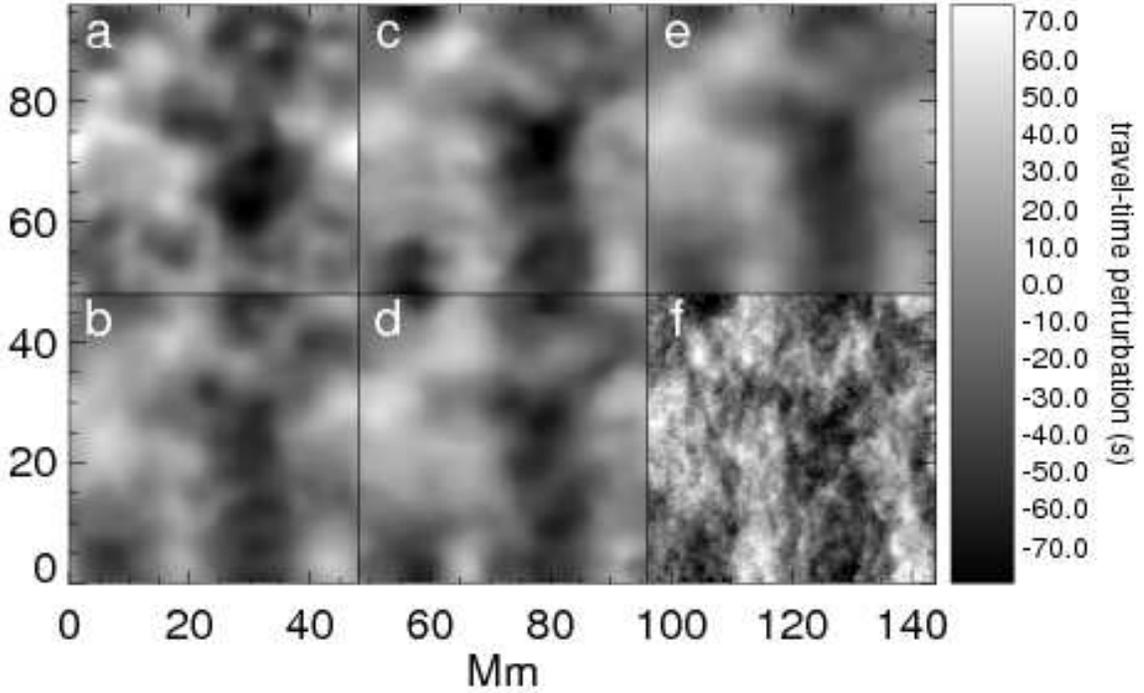}
\caption{
Maps of the $p$-mode travel-time perturbations 
determined from lateral-vantage HH using the east/west
quadrant pair and
a focus depth of 0.7 Mm below the simulation surface for
the following frequency bandpasses:
a) 2.5 -- 3.5 mHz, b) 3.5--4.5 mHz, c) 4.5--5.5 mHz,
d) 5.5-6.5 mHz, e) the average of all four frequency
bandpasses, f) the time average of the x component of the horizontal velocity,
$v_x$, at a depth of 0.7 Mm. The color bar gives the units of
the travel-time perturbation for panels a--e, while
the values for the velocity (panel f) range from 1.1 km/s (black) 
to -1.1 km/s (white). A positive value of $v_x$ 
corresponds to a flow towards the right.
}
\label{dt_freqs}
\end{figure}

An examination of the observed travel-time perturbation maps reveals a 
fair degree of correlation amongst different frequency 
bandpasses for a given focus depth, particularly for the
shallower depths ($\leq$ 5 Mm)  selected for analysis. 
An example, at the 0.7 Mm focus depth,
is shown in Figure~\ref{dt_freqs}.
Correlation coefficients between perturbation maps 
at different frequency were computed after applying 
Gaussian smoothing with varying FWHM
to the maps. As expected, the correlations
improve with increased smoothing. With a Gaussian FWHM of 4 Mm,
the correlation coefficients are of the order of 0.7 for the
shallowest depths  and decrease to about 0.5 at a depth of 6 Mm.
At depths of 7 and 8 Mm, the 1 mHz bandpass maps are essentially
uncorrelated with each other.
As with analogous travel-time observations in the Sun, the measurement
noise from these simulations is most likely due to realization noise
caused by stochastic excitation of the $p$-modes. In the
solar case, the properties of the measurement noise have
been extensively explored \citep[e.g.][]{Gizon2004}.

At depths where comparisons between frequencies
are meaningful, we do not detect any clear systematic variation
of the travel-time perturbations with frequency. Typically, the
slope of a linear fit of the travel times
between any two frequencies differs from unity by less than
10\%. A somewhat surprising result is that 
the variance of the 1~mHz bandpass maps, defined
as the average over all pixels of the squared difference between
each map and an average of maps over all four bandpasses (hereafter
referred to as a frequency-averaged map),
is essentially the same for all frequencies between 3 and 6 mHz.

As expected, there is a good correspondence, particularly 
at shallower depths, between the travel-time maps and the relevant component
of the horizontal velocity. This is particularly evident in comparisons
between the frequency-averaged perturbation maps and the simulation flows
(e.g.\ compare Figure~\ref{dt_freqs}e and Figure~\ref{dt_freqs}f).
However, it is also apparent that the ratio of the magnitude of the 
travel-time perturbations to their values at the shallowest focus depth
(shown by the dotted lines in Figure~\ref{ratio}) falls off considerably 
faster with focus depth than a similar ratio of the simulation 
flow magnitudes (solid lines).

\subsection{Comparisons with Model Travel-time Perturbations} \label{obs-mod}

\begin{figure}[htbp]
\epsscale{1.}
\plotone{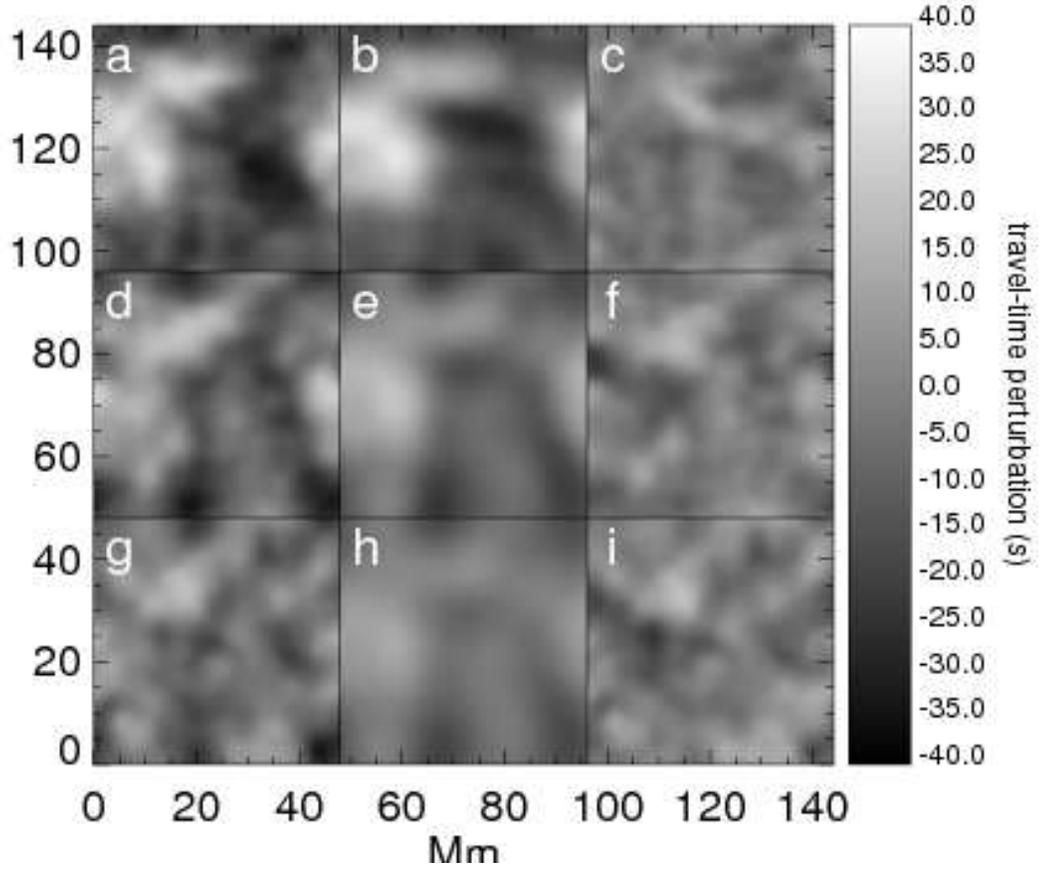}
\caption{
Comparisons of the observed $p$-mode travel-time perturbations 
determined from lateral-vantage HH with the perturbations
computed from a forward model for several focus depths:
a) the observed,  b) model, and c) residual (observed - model) 
perturbations at a focus depth of 3 Mm, d-f) the corresponding 
perturbations at a focus
depth of 5 Mm, g-i) the corresponding perturbations
at a focus depth of 7 Mm below the surface
}
\label{obs-model}
\end{figure}

Using the methods described in \S~\ref{models} we derive
the model travel-time perturbations ($\delta\tau_{\rm mod}$) 
for comparisons with
the observed values. Because of the restriction to low frequency modes
described in \S~\ref{models} we compute $\delta\tau_{\rm mod}$ only 
for the lowest frequency bandpass (${\nu}_0 = $ 3 mHz).  
Figure~\ref{obs-model} shows a comparison between the frequency-averaged 
travel-time perturbations and the 3-mHz model
for three focus depths; 3, 5, and 7 Mm below the surface. Also
shown are the residuals after subtracting the model perturbations
from the observations. There is good agreement between
the model and observations for the shallower depths.
At greater depths the similarity becomes less apparent. 
The patchy appearance of the residual maps is likely due
to the properties of the realization noise which is coherent
over distances of about half of a $p$-mode wavelength \citep{Gizon2004}.
The similarity of the residuals at different focus depths 
follows from the highly overlapping set of modes used in
the analysis for the different depths (see Table~\ref{tbl-1}).
We define the square of the deviation from the model, $\sigma_{\rm mod}^2$, 
of each
observed travel-time perturbation map ($\delta\tau_{\rm obs}$) 
as: 

\begin{equation}
\sigma_{\rm mod}^2 =  \frac{1}{N} \sum (\delta\tau_{\rm obs} - \delta\tau_{\rm mod})^2 , \label{eq:sigma_mod}
\end{equation}
where the summation is over all $N$ pixels in the maps.

\begin{figure}[htbp]
\epsscale{1.}
\plotone{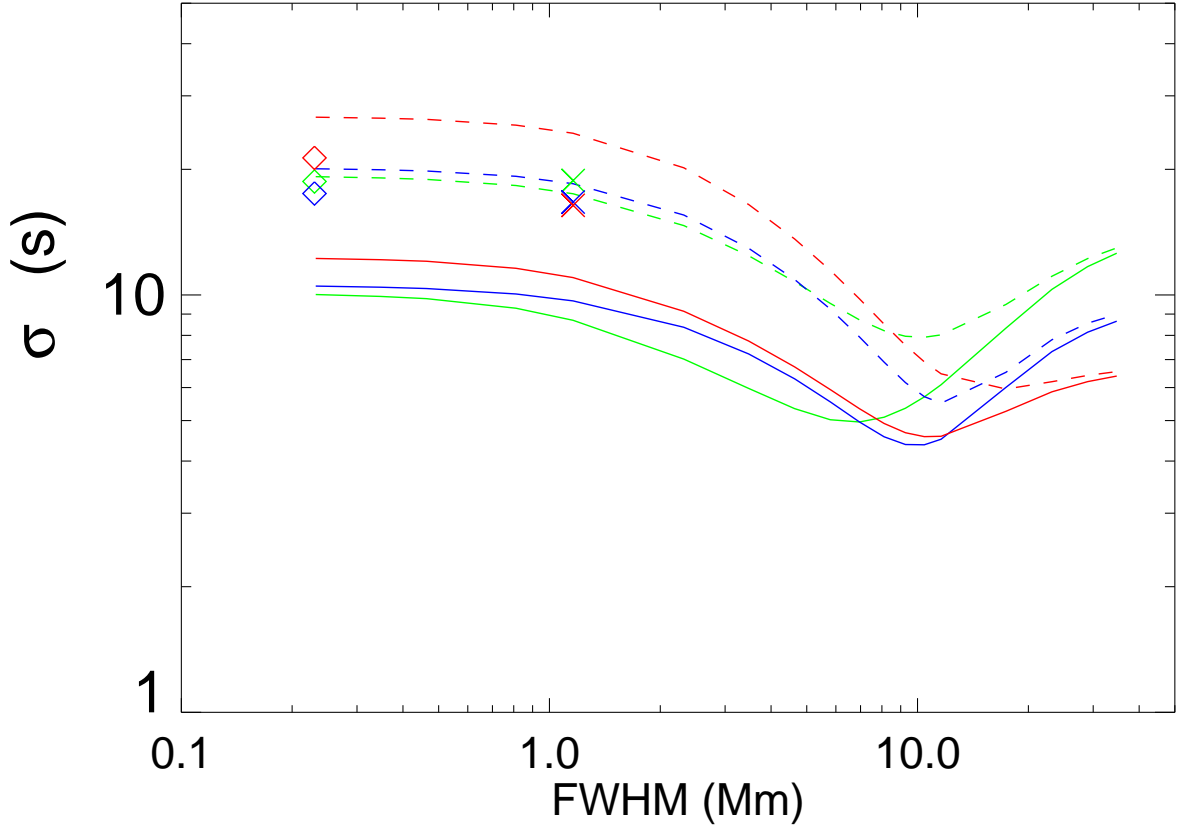}
\caption{
Measurements of the deviation from the forward models of
the travel-time perturbations as functions
of the degree of spatial smoothing of the observed
perturbations. The FWHM indicates the full-width-at-half-maximum
of the two dimensional Gaussian used to smooth the travel-time
maps.
The dashed lines indicate the results
using  1-mHz wide frequency filters centered at 3 mHz
for focus depths of 3 Mm (green), 5 Mm (blue), and 7 Mm 
(red) below the surface. The solid lines show the
results for the frequency-averaged measurements.
The diamonds indicate the standard deviations
of the travel-time perturbations across the four
independent frequency filters. The X symbols indicate
similar standard deviations (across only three
filters) for MDI observations (see text). The colors of the symbols
indicate the same focus depths as the lines.
}
\label{noise}
\end{figure}

To see how $\sigma_{\rm mod}$ depends on smoothing, we spatially
convolve each observed perturbation map by two-dimensional
Gaussian functions with varying
FWHM. Figure~\ref{noise} shows the resulting $\sigma_{\rm mod}$ as a function
of FWHM. The different colors indicate 
different focus depths, and the results are
shown for both the 3 mHz bandpass (dotted lines) and the
average over the four frequency bandpasses (solid lines).
The deviation in the frequency-averaged perturbation maps is reduced
from the single 3 mHz bandpass
by a factor of two (i.e.\ the decrease from the dotted to
solid lines) as expected from independent measurements
with similar uncertainties (although this improvement
declines with more smoothing). 

If each pixel represented an independent sample of the travel-time perturbation
with an uncertainty which is normally distributed,
we would expect $\sigma_{\rm mod}$ to decrease with the inverse of the
FWHM. Figure~\ref{noise} shows, however, a considerably weaker decrease of the 
deviation, most likely due to the finite wavelengths of the modes used to
produce the maps \citep{Gizon2004}. In other words, the maps are already ``smoothed'' by 
finite-wavelength effects which can be readily seen in the raw perturbation maps.
With values of FWHM more than about 10 Mm, the effects of smoothing increase
the departure from the model signatures, since the level of smoothing 
compromises the ability to resolve the dominant
flow structures (e.g.\ Figure~\ref{cells}).
When ``over-smoothed'' in this way, there is less 
(and sometimes no) improvement achieved by frequency averaging.
The values of $\sigma_{\rm mod}$  are generally
similar for all of the depths shown, although the deeper two measurements
can apparently withstand slightly greater smoothing 
to obtain smaller deviations from the model.

We have examined the statistics 
of the travel-time measurements in the independent one mHz-wide frequency 
bandpasses by computing the standard deviation of the measurements in 
each pixel from the mean over the four frequency bandpasses. 
Some caution is warranted in 
directly comparing the pixel-average of this quantity, which we 
denote $\sigma_{\nu}$,  
with $\sigma_{\rm mod}$ when the data is smoothed. 
For example, substantial smoothing can reduce the
difference between frequency bandpasses to an arbitrarily 
small value, even while
both differ substantially from the model. 
A correspondence 
between these two quantities is expected however, in the limit of little or no
smoothing, if there are no systematic differences between frequency
bandpasses of the perturbations or their uncertainties. 
The values of $\sigma_{\nu}$ from 
the maps of $\delta\tau_{\rm obs}$, smoothed with a Gaussian FWHM 
only minimally larger than the pixel size, are indicated
by the diamonds in Fig.~\ref{noise}.  
The correspondence is particularly striking for depths of 3 and 5 Mm. 
At a depth of 7 Mm, the ratio $\sigma_{\nu} / \sigma_{\rm mod}$ is about 0.8. 

\subsection{Comparisons with MDI Observations} \label{mdi}

\begin{figure}[htbp]
\epsscale{1.}
\plotone{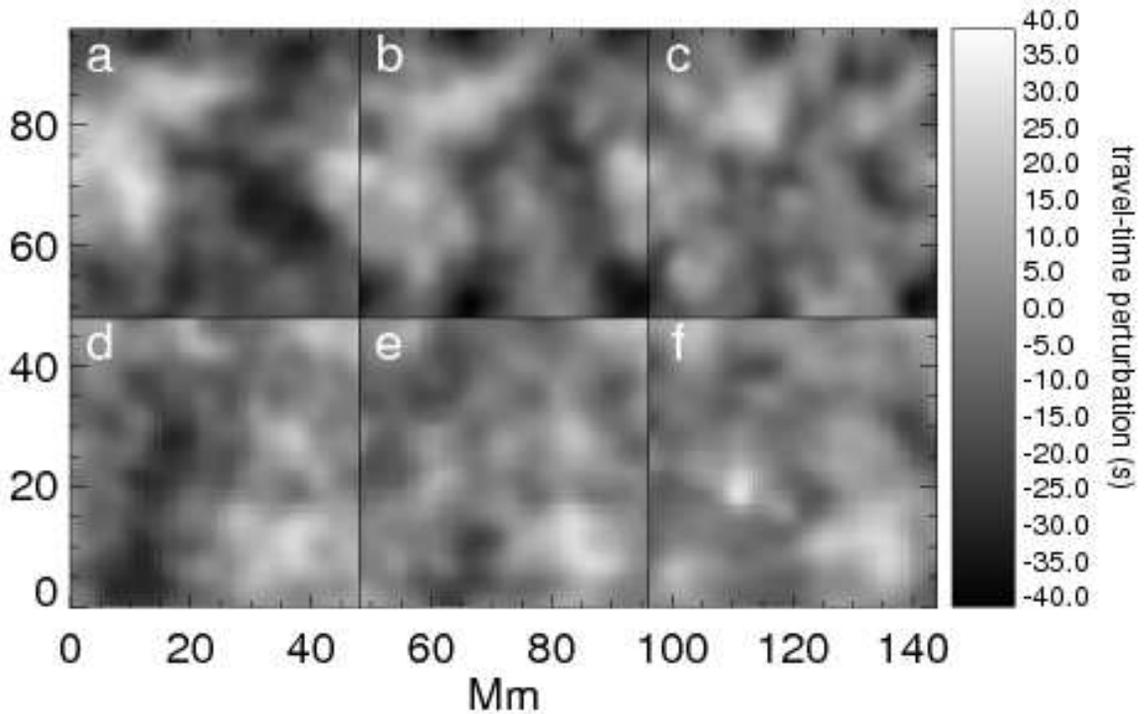}
\caption{
Maps of frequency-averaged travel-time perturbations
measured from the simulations and from a cropped
region of the Sun observed with {\it SOHO}/MDI. 
The top three panels show the results for
the simulations at focus depths of a) 3~Mm, b) 5~Mm, and c) 7~Mm,
and the bottom three panels show the same depths for the MDI
observations. Only frequency bandpasses centered at 3, 4, and
5 mHz were used to make the simulation and MDI averages 
displayed here.
}
\label{MDI}
\end{figure}

It is of interest to compare the properties of the simulation 
observations with results obtained with actual solar observations.
An 8.53 hour time series, with 60 second cadence, of solar
images was processed identically to the simulated data
reported here. The data set consists of 512 Dopplergrams obtained
by {\it SOHO}/MDI \citep{Scherrer1995} on 1999 June 1.
The MDI Dopplergrams were interpolated onto Postel's projections,
tracked with a Carrington rotation rate and centered on longitude
$L=306^\circ$ and latitude $B=7^\circ$. The projected frames consist
of a grid of 512 by 512 pixels with a resolution of 828 km per
pixel, obtained after a two by two pixel averaging of the original high
resolution pixels. Travel-time perturbations were computed,
with the same HH analysis as the simulations, 
for focus depths of 3, 5, and 7 Mm below the surface over
1-mHz temporal bandpasses centered at 3, 4, and 5 mHz.
In general, HH travel-time measurements from solar data, unlike
the simulations, are not reliable at higher frequencies.
Figure~\ref{MDI} shows cropped (48 by 48 Mm) travel-time
perturbation maps, averaged over all frequency bandpasses,
of a region close to the center of the solar disk,
compared with maps obtained from the simulated data at the
same focus depths. It would be difficult to identify, from
general appearances alone, which measurements came from
the artificial data and which came from solar data 
(except perhaps for the finer pixel size noticeable in the simulations). 

To assess the noise in the MDI observations, we compute
the pixel-averaged standard deviation across the three
frequency bandpasses. For these measurements, only
a minimal smoothing (FWHM $\approx$ 1.2 Mm) is applied.
The results are shown by the X symbols in Figure~\ref{noise}.
These values agree very well with the single-bandpass model-deviations
of the simulation measurements after smoothing
with the same Gaussian function. We conclude that
the magnitude of the uncertainties in HH travel-time perturbation
measurements made with 8 hr of solar observations are 
very similar to those obtained with the simulated data
analyzed here.

\subsection{Depth Contribution of Signal} \label{summary}

We use the forward models to
examine the relative contribution with depth
to the expected travel-time perturbations from
the flow structures. To successfully infer
the magnitude of subsurface flows from the
travel-time observations, we will very likely
require that the expected perturbations due to
flows at a desired focus depth exceed the 
noise present in the observations. Thus we need
to extract the relevant portion of the expected
signatures (i.e.\ due to the flow near the target focus) 
from the total travel-time perturbations (which result 
from flows over the entire depth range sampled
by the modes used in the measurements).

Assessing the contribution to the travel-time
signatures of flows near the surface are of
particular interest. This is motivated by
the desire to understanding the rapid
decrease in the observed travel-time perturbations
with focus depth as exhibited in Figure~\ref{ratio}.
Previous HH analyses and modeling of supergranulation observed
with {\it SOHO}/MDI have indicated the possibility
of a rapid decrease, and eventual reversal, of the
travel-time perturbations with depth \citep{Braun2003, 
Braun2004}. An explanation was offered by \cite{Braun2003}
that the results were consistent with an increase with focus depth
of the near-surface contribution to the travel-time perturbations
from oppositely directed flows in neighboring supergranule
cells. These near-surface contributions increased with focus
depth because the pupils needed for lateral-vantage HH increase
in radii with depth (e.g.\ Table~\ref{tbl-1}) eventually
exceeding the size of the supergranular cells.
Forward models assuming simple, shallow, cellular flows
were able to reproduce this effect \citep{Braun2004}, and
it is of considerable interest to see if such behavior
exists for the flow patterns present in the simulations
of \cite{Benson2006}.

To assess the depth-dependence of the contribution to
the travel-times due to subsurface flows we perform a horizontal
convolution of the kernels (\S~\ref{models}) 
with the $x$ component of the flow fields. 
Figure~\ref{slice.0043} shows a vertical slice of the $x$ component of 
the flows, a slice through the kernel functions (for a focus
depth of 3~Mm), and a slice through the horizontal convolution of the 
two.  Figs~\ref{slice.0072} and \ref{slice.010} show
the the same for focus depths of 5 and 7~Mm, respectively. 
What is readily apparent in these figures is that, while the
flows fall off gradually with depth (Figure~\ref{slice.0043}a), 
the contribution to the travel-time perturbations 
(Figs~\ref{slice.0043}c, \ref{slice.0072}c, and  \ref{slice.010}c) 
due to near-surface layers (with depths from 0 - 1 Mm) is 
disproportionately large. This
results from the strong near-surface sensitivity of
the travel-time kernels 
(Figs~\ref{slice.0043}b, \ref{slice.0072}b, and  \ref{slice.010}b).
It is noteworthy that the root-mean-square (RMS) of
the travel-time perturbation integrated from the surface
to a depth of 2 Mm is, for all three depths shown, roughly
the same as the RMS of the perturbation integrated from 2 Mm to
the bottom of the simulation, where the RMS is computed
over the full horizontal domain.

For what follows we define the ``target signal'' to
be the contribution to the travel-time perturbation
over a ``target-range'' of depths
about a given focus depth.  This range is intended to
be proxy for the depth resolution of 
any modeling of the observations 
(through either forward or inverse methods).
For a depth of 3 Mm below
the surface, the integration is carried out from 2 Mm to
6 Mm. For the focus of 5 Mm, the target range is from 4 Mm to 8 Mm
and for 7 Mm, the target range used is 5 to 10 Mm. These ranges
were selected by a visual inspection of the sensitivity
functions (e.g.\ Figs~\ref{slice.0043}b, 
\ref{slice.0072}b, and \ref{slice.010}b). 
Figs~\ref{slice.0043}d, \ref{slice.0072}d, and  \ref{slice.010}d
show plots of the total travel-time perturbations (blue lines),
the target signal (red lines), and the 
contribution away from the target (black lines), over a slice
of the data cube. Note that Figs.\ \ref{slice.0072} and \ref{slice.010}
indicate a positive contribution to the travel-time
perturbations at $x \approx 30$ Mm from the
near surface, which is opposite
in sign to what is expected for the rightward-directed
flow at this location (Figure~\ref{slice.0043}a). 
This contribution is due to flows located in the
near surface lobes of the kernel which have opposite
sign from the flow at the focus. These measurements
offer support to the suggestion that the rapid decrease 
(and potential reversal in sign) of the total travel-time 
perturbation with increasing depth results from the
near-surface contribution of nearby (and oppositely directed) 
flows.

To estimate the signal-to-noise ratio (SNR) required
for the detection and modeling of the subsurface flows
in the simulations we use the RMS of the
target signals as defined above.
For the noise levels,  we use the optimal (i.e. minimum) values 
of $\sigma_{\rm mod}$ from Figure~\ref{noise} for each depth.
For 3 Mm, this is 5 seconds, and for 5 and 7 Mm, this is
4.5 seconds.  Table~\ref{tbl-2} shows the values of
the RMS for the total travel-time perturbation (column 2), the RMS of the 
target signal (column 3) and the resulting SNR (column 4) for the three 
depths listed in column 1. These results do not substantially
change if the ``target'' is adjusted to include flows at
deeper layers.  For example, extending the lower depth limit of each
target range to the bottom of the simulation
box increases the RMS target signal and the resulting SNR
by approximately 5--15\% for the depths listed in Table~\ref{tbl-2}. 
It is clear that the SNR is too small for 
detecting subsurface flows in these simulations 
at and below 5 Mm with 8 hr of data.
Note that a SNR of one implies the observed travel times are 
consistent, within the noise, with no flows. 
A SNR greater than unity is required 
for an unambiguous detection of a flow.

\section{Discussion}            \label{discussion}

A major finding of this work is that 
lateral-vantage holography recovers the model
travel times about as well as one should expect.
In other words, the RMS of the residuals (observed
minus model) is very close to the noise levels
as deduced by the variance between
independent frequency bandpasses.
We can also infer from the similarity of
results across the different bandpasses that the 
validation of the method applies across a wide
range of temporal frequencies. In addition,
we have gained a fair degree of confidence in the
use of the Born approximation for modeling
$p$-mode travel-times for flows similar to the
the ones simulated here. We note, however, that
models have only been computed in one frequency 
bandpass and that further effort is needed to
extend the modeling to higher frequencies. One important
issue not addressed here is the potential
improvement in the spatial resolution of 
flows (or other perturbations) deduced with observations made
with increasing temporal frequency of the $p$-modes
employed in the analysis. This is plausible if
the resolution scales with the mode wavelength
which, for a mode propagating through a fixed
focus, varies inversely with the temporal frequency.

\begin{figure}[htbp]
\epsscale{1.}
\plotone{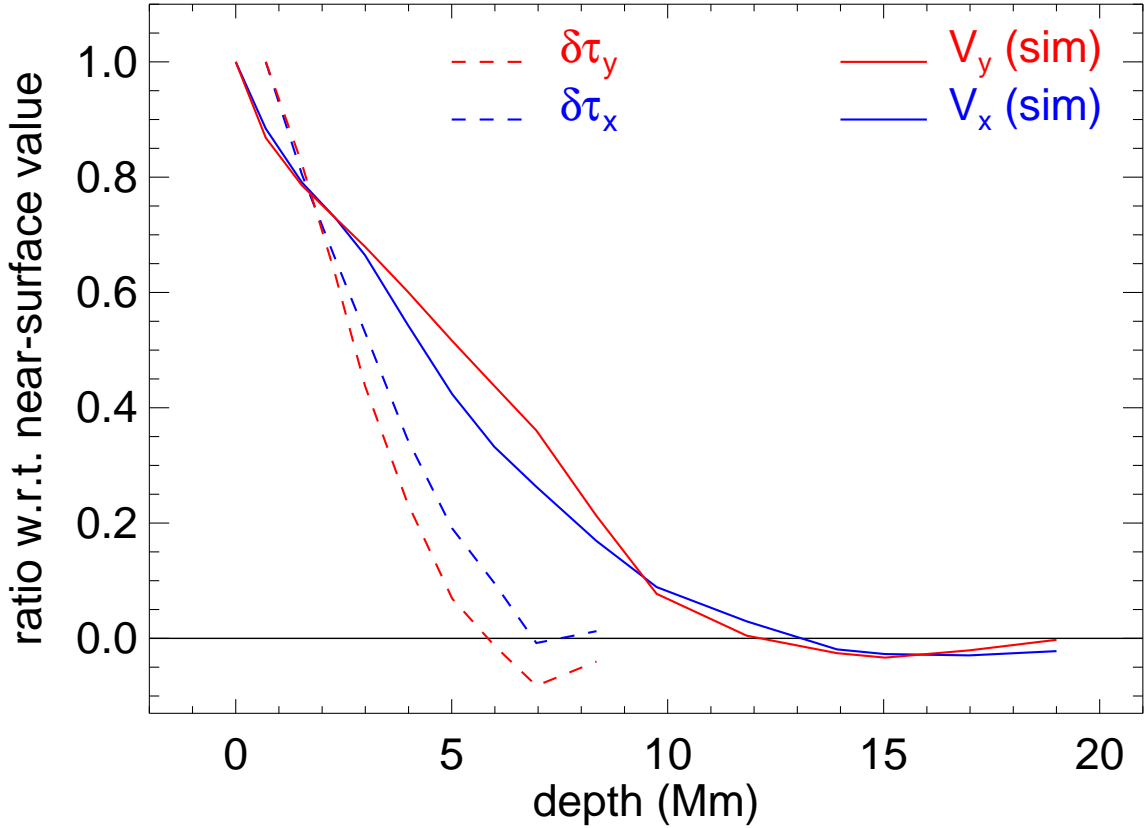}
\caption{The ratio with respect to near-surface values of
the simulated horizontal flows (solid lines) and the 
measured holographic travel-time perturbations (dotted lines) as a function
of depth. The
blue and red lines indicate the measurements for the
$x$ and $y$ components of each quantity respectively.
For the travel time perturbations, the depth corresponds
to the depth of the focus in the lateral vantage.
For both flows and travel time perturbations, the ratio is
defined to be the slope of a linear least-squares fit between the 
relevant quantity at the indicated depth with the quantity evaluated
at a ``near-surface'' depth. For the simulated flows this depth
is the surface of the simulations. For the travel-time perturbations 
this depth is 0.7 Mm below the surface.
}
\label{ratio}
\end{figure}

\begin{figure}[htbp]
\epsscale{1.}
\plotone{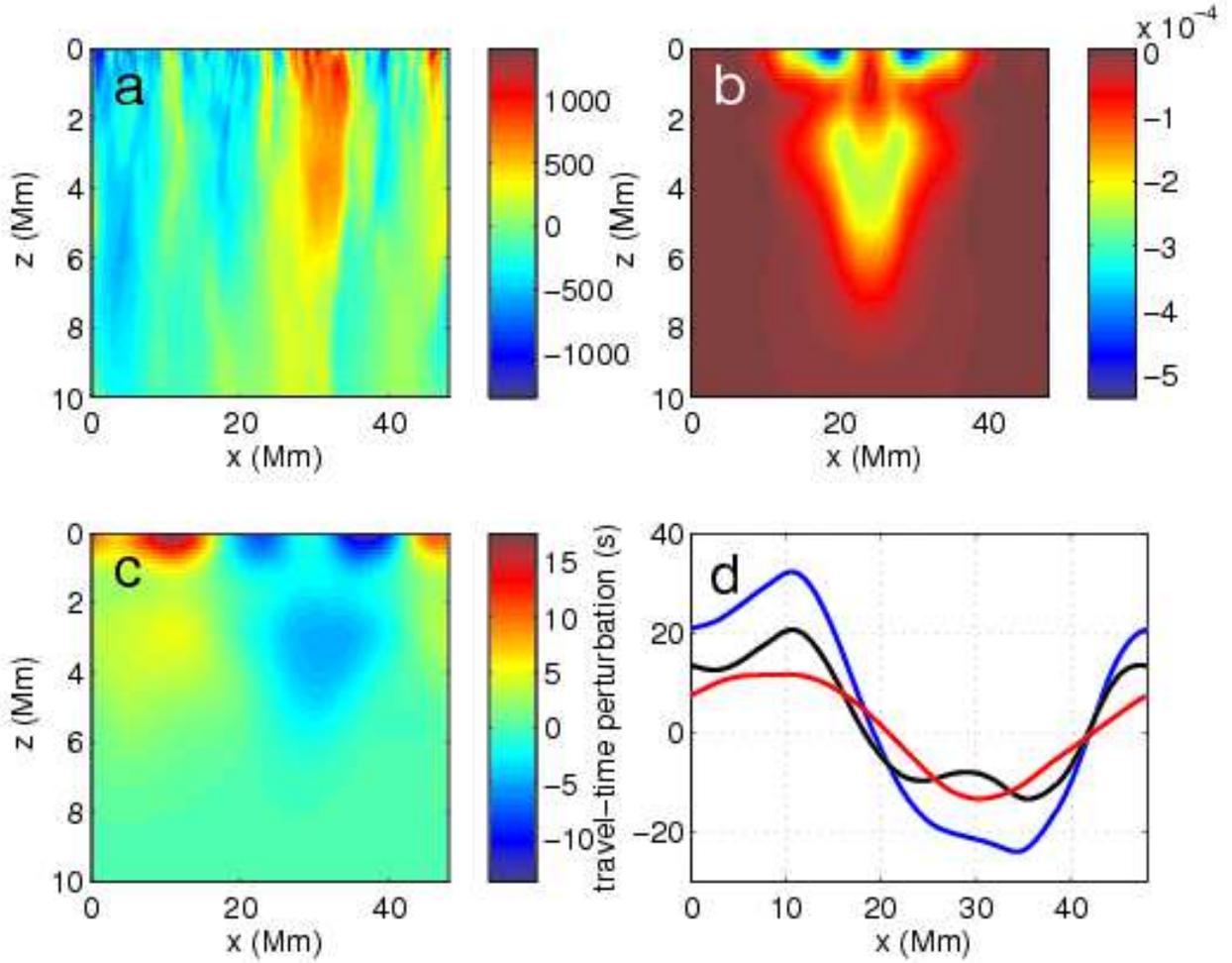}
\caption{
a) a vertical slice at $y=24$~Mm through the time-averaged 
$x$ component of the flow from the simulation.  
b) a slice through the travel-time kernel for a focus depth 
of 3~Mm, the color shows values in units of s Mm$^{-3}$ /(m/s).   
The travel-time kernel is defined as ${\bf K} /2\pi{\nu}_0$, where
${\bf K}$ is given by
eq.~\ref{eq:kernel} for the 
antisymmetric phase shift computed between the east and west
pupil quadrants,
and ${\nu}_0 = $ 3 mHz. The travel-time kernel is negative and
relates a positive (negative) flow in the $x$ direction to a
travel-time decrease (increase).
c) a vertical slice, at $y=24$~Mm, through the horizontal 
convolution of the kernel with the $x$ component of the flow field; 
the color bar has units of s Mm$^{-1}$.   
d) the total travel-time perturbation (blue), the target signal (red)
and the remaining (total minus target) signal (black) 
as functions of the $x$ coordinate of the focus position.
The target signal is integrated over a range of depth from 
2 to 6 Mm.
}
\label{slice.0043}
\end{figure}

\begin{figure}[htbp]
\epsscale{1.}
\plotone{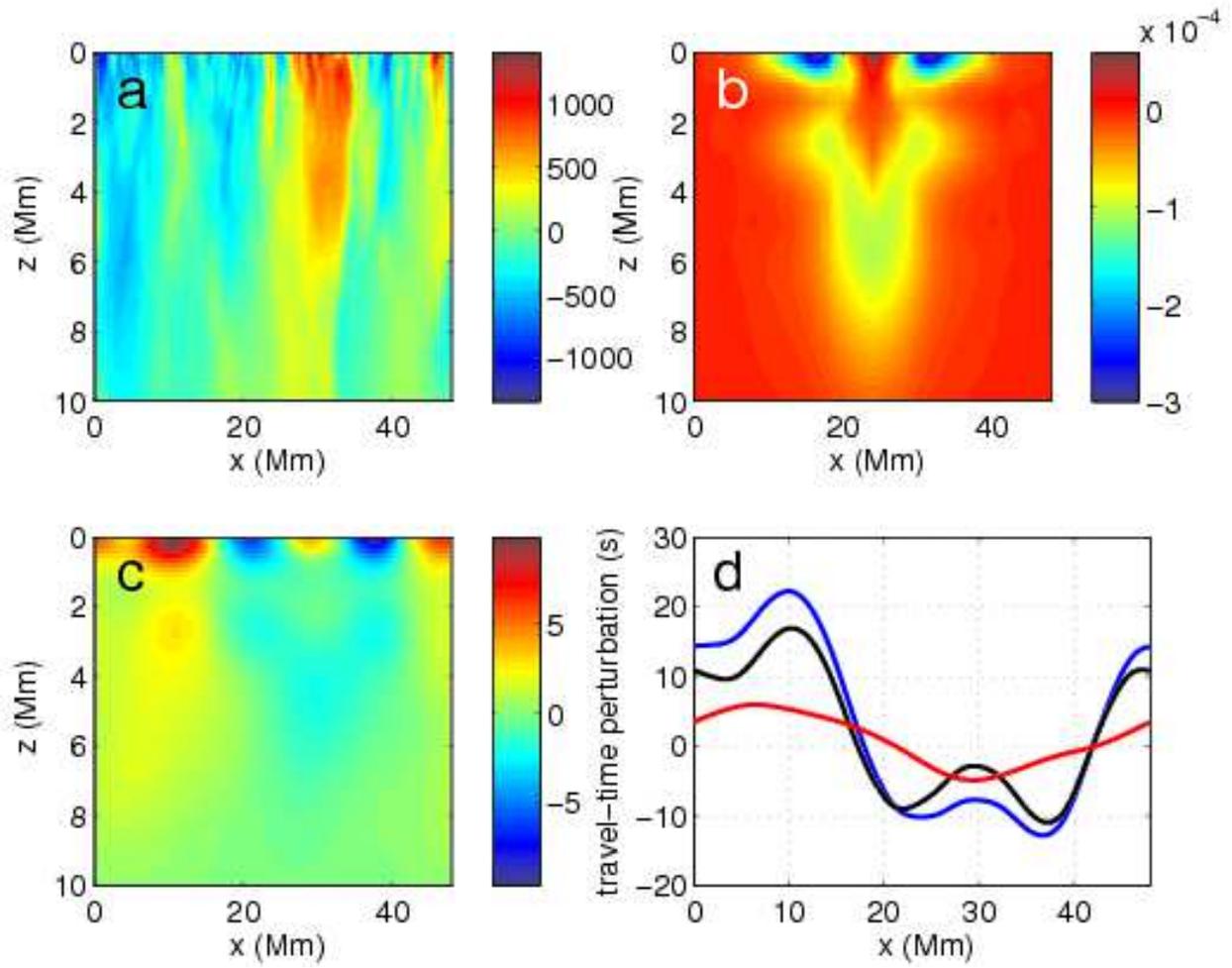}
\caption{
Same as Figure~\ref{slice.0043} except now the focus depth is 5~Mm. 
The target signal is integrated over a range of depth from 
4 to 8 Mm.
}
\label{slice.0072}
\end{figure}

\begin{figure}[htbp]
\epsscale{1.}
\plotone{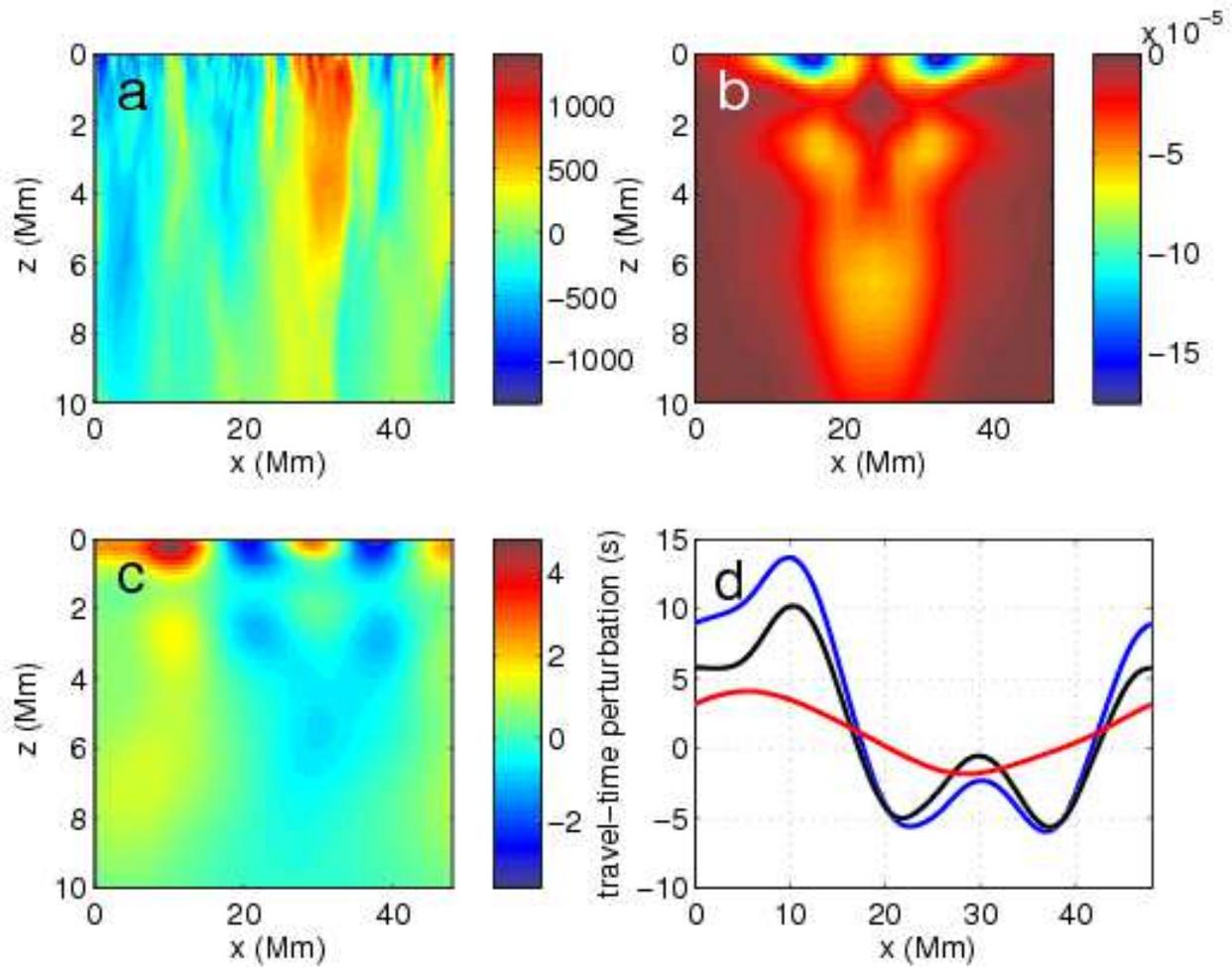}
\caption{
Same as figure Figure~\ref{slice.0043} except now the focus depth is 7~Mm. 
The target signal is integrated over a range of depth from 
5 to 10 Mm.
In panel c, at $x\approx 30$~Mm notice the positive contribution to the 
travel-time perturbation from the near-surface.  
This contribution is due to flows located in the near-surface 
lobes of the kernel and with opposite sign to the flow at the focus point. 
}
\label{slice.010}
\end{figure}

We have made a quantitative
assessment of the contribution of near-surface layers
to travel-time measurements as a whole.
A major consequence of our result is a plausible
explanation for the relatively rapid decrease 
with depth (and eventual sign reversal) of the travel time
perturbations, which has also been detected in MDI observations of
solar supergranulation. This trend is not necessarily detrimental to
successful modeling of subsurface flows, given
a sufficient SNR in the travel time measurement and
the assumption that the near-surface perturbations 
are well understood.  Most modeling efforts typically
include an assessment of flows at a range of depths, including
the near-surface regions. 
In the simulations performed here, the horizontal flows
decrease gradually with depth (Figs~\ref{ratio} and \ref{slice.0043}a).
However, strong shear layers immediately below the
photosphere or the existence of surface perturbations due to 
magnetic fields (neither of which is present in these
simulations) could enhance the near-surface
contributions and potentially complicate the modeling
of flows.  Some possible complications include strong surface
perturbations which are not easily modeled under assumption of the Born 
approximation, or that might require the use of ``surface terms''
which are not routinely employed in local helioseismology
\citep[see][]{Braun2006}. Numerical simulations provide
an excellent means of testing these possibilities.

In as far as these simulations predict 
the types of flows likely to be present in Sun, it is appropriate 
to extrapolate our results to evaluate the prospects
for the detection and modeling of actual solar flows.
We assume similar noise levels, as confirmed by the
comparisons performed in \S~\ref{mdi}.
An increase in the temporal duration of the observations
from 8 to 24 hrs of observations, or roughly the lifetime
of the supergranules, should produce an increase in the SNR
by a factor of $\sqrt 3$. On the other hand, our experience
has shown that only the 3, 4, and 5 mHz frequency bandpasses
can be used successfully in solar measurements, so that the nominal
increase in SNR is a more modest factor of about 1.5. The
resulting SNR values are shown in column 5 of Table~\ref{tbl-2}.
The general result, that supergranule-sized flows are essentially 
undetectable using current methods below depths around 5 Mm,
is not substantially changed.

It is fair to point out that our pessimistic
conclusions about detecting small-scale subsurface flows 
below depths of 5 Mm are based
on idealized conditions and assumptions. To the extent
that these conditions may not be representative of
actual or proposed solar measurements and modeling
it is plausible that our conclusions may not be fully relevant.
On the other hand it has been our goal to err, if at all, 
on the side of optimism in computing the SNR values quoted
here. Our noise estimates are inferred from
forward models constructed directly from the subsurface flows present
in the simulations, and thus
represent an idealized, ``best-case,'' scenario. 
Thus, the SNR values quoted do not address possible
errors introduced by modeling limitations or 
systematic effects encountered in solar measurements.
\cite{Zhao2007} have performed inversions
of time-distance measurements performed on this simulation and find
only weak correlations between the actual and inverted
flows at depths 3 Mm and greater below the surface. This
appears to be consistent with our findings regarding 
SNR estimates, although the methods employed are different.

While the use of lateral-vantage HH
is a fairly efficient means of measuring the effects
of the horizontal component of the flow field (see \S~\ref{analysis}),
we recognize that additional information may be used to
infer the properties of subsurface flows. These include
using helioseismic measurements of vertical flows and the assumption
of mass conservation as an additional constraint on
horizontal flow components, or matching subsurface flows to 
those observed at the surface through other means 
(e.g.\ direct Doppler measurements 
or feature-tracking). The degree to which this additional information
can be used to substantially improve our ability to measure flows deeper than 
a few Mm below the photosphere, however, remains to be determined.
We also recognize that solar supergranulation has a
spatial scale somewhat larger than the
cells explored in these simulations, which may permit
somewhat more spatial smearing and subsequent reduction
in noise.  Some improvement might be expected by
increasing the temporal duration of the observations, since
some supergranules last as long as several days. And
obviously, many subsurface dynamic phenomena, both known
and unknown, will still be accessible after temporal and
or spatial averaging. It is also important to keep in mind
the spatial and temporal scales of the flows for which our findings
are relevant. Clearly, SNR values far greater than unity
are possible, even at substantial depths below the photosphere,
for measurements of large-scale or long-lived structures such
as meridional and zonal flows.

\cite{Birch2006} presented a method to combine
travel-time measurements, from time-distance helioseismology,
of thousands of supergranules, allowing the exploration
of supergranule-averaged flows.  In general, 
a parameterization of small-scale flows
(in terms of statistical moments or 
Fourier components, for example) 
might be more amenable to modeling than the
reconstruction of individual flow vectors 
\citep[e.g.][]{Woodard2006, Woodard2007}.

Forward modeling efforts with both solar and artificial
data appear to offer considerable utility for understanding
both advantages and limitations of helioseismic methods.
It would be useful to extend the analysis presented here 
to address a variety of additional issues. Understanding
the trade-off between maximizing spatial resolution in 
modeling perturbations and minimizing the uncertainties in the
models is a major issue which can be explored through
forward modeling. For example, it should be 
straightforward to measure the signal and noise values for 
different spectral components (in the spatial domain) of flows and
other perturbations to address this. 

Finally, we emphasize the importance of
numerical simulations for helioseismic analysis. Some improvements 
are now becoming available or are under preparation. These include
simulations which cover larger areas and extend deeper into the
solar interior and are carried out in spherical geometry.
There is also a need to compliment realistic simulations (which give
the best current estimate of what's likely to be found in the
solar interior) with the ability to prescribe
and adjust perturbations (in order to fully test observing and modeling
methods and to allow for surprises in the Sun). 
The addition of magnetic
fields to numerical simulations will be of particular utility to
test both their influence on convective and other flows, and
their influence on local helioseismic
measurements.

\acknowledgments
DCB and ACB are supported by funding through NASA contracts NNH05CC76C and
NNH04CC05C, NSF grant AST-0406225, and a subcontract through the 
HMI project at Stanford University awarded to NWRA. The numerical 
simulations were carried out with support by
NASA grant NNG04GB92G and NSF grant AST-0605738.

\begin{deluxetable}{ccccc}
\tablecolumns{3}
\tablewidth{0pc}
\tablecaption{Pupil radii.\label{tbl-1}}
\tablehead{
\colhead{Depth} & \colhead{Pupil radii} & \colhead{$\ell$ @ 4mHz}\\ 
\colhead{(Mm)} & \colhead{(Mm)} & \colhead{ }\\ 
}
\startdata
0.7 &  0.7 - 11.1 & 980 - 1370 \\
1.5 &  1.2 - 12.5 & 850 - 1190 \\
2.3 &  1.8 - 13.9 & 750 - 1060 \\
3.0 &  2.1 - 14.6 & 680 - 960 \\
4.0 &  2.8 - 16.7 & 590 - 840 \\
5.0 &  3.5 - 18.8 & 530 - 740 \\
6.0 &  3.5 - 24.4 & 470 - 670 \\
7.0 &  4.2 - 29.9 & 420 - 600 \\
8.3 &  4.9 - 39.0 & 370 - 520 \\
\enddata
\end{deluxetable}

\clearpage

\begin{deluxetable}{ccccc}
\tablecolumns{5}
\tablewidth{0pc}
\tablecaption{Signal-to-Noise Estimates\label{tbl-2}}
\tablehead{
\colhead{Depth} & \colhead{RMS total signal} & \colhead{RMS target}  & \colhead{SNR} & \colhead{SNR}\\ 
\colhead{(Mm)} & \colhead{(s)} & \colhead{(s)} & \colhead{8 hr sims.} & \colhead{ 24 hr Sun}\\ 
}
\startdata
3.0 & 15.6 & 8.7 &  1.7 & 2.6 \\
5.0 & 10.5 & 3.6 &  0.8 & 1.2  \\
7.0 & 7.4 & 1.7 &  0.4 & 0.6  \\
\enddata
\end{deluxetable}

\clearpage

\end{document}